%Paper: hep-th/9209063
%From: TEMPLE@vax2.concordia.ca
%Date: Thu, 17 Sep 1992 13:17 EDT

\magnification\magstep1
\settabs 6\columns
\vskip 1in
\centerline{\bf Differential cross-sections and escape plots for}
\centerline{\bf low energy $SU(2)$ BPS magnetic monopole dynamics}
\vskip 0.5in
\centerline{M. Temple-Raston, D. Alexander}
\vskip 0.2in
\centerline{Department of Mathematics and Statistics,}
\vskip 0.1in
\centerline {Concordia University, 7141 Sherbrooke Street West,}
\vskip 0.1in
\centerline {Montr\'eal, Qu\'ebec H4B 1R6}
\vskip 0.5in
\noindent
{\bf Abstract}
\smallskip
\noindent
We compute the low-energy classical differential scattering cross-section
for BPS $SU(2)$ magnetic monopoles using the geodesic approximation to the
actual dynamics and 16K parallel processors on a CM2.  Numerical experiments
suggest that the quantum BPS magnetic monopole differential cross-section
is well-approximated by the classical BPS magnetic monopole differential
cross-section.  In particular, the expected quantum interference effects
for bosons at scattering angle $\theta=\pi/2$ (CoM frame) are contradicted
numerically.  We argue that this is due to the topology of the classical
configuration space for these solitons.  We also study the scattering and
bounded classical motions of BPS dyons and their global structure in phase
space by constructing `escape plots'.  The escape plots contain a surprising
amount of structure, and suggest that the classical dynamics of two BPS
$SU(2)$ magnetic monopoles is chaotic and that there are closed and
bounded two dyon motions with isolated energies.
\vskip 0.25in
\noindent
\+MR subject classification: 70F05, 58F13, 81Gxx\cr
\+e-mail: temple@vax2.concordia.ca\cr
\vfill\eject
\openup 2\jot
\noindent
{\bf Introduction}
\medskip
\noindent
Determining the dynamics of solitons in mathematical physics is a
particularly difficult problem because it involves solving hyperbolic
partial differential equations (PDEs).  As a result, numerical techniques
have been employed to numerically `solve' the equations.  Unfortunately,
the effectiveness of numerical schemes for PDEs diminishes as the dimension
of the space on which the PDEs are defined, increases.  Therefore, to
extend our understanding of soliton dynamics into areas for which the
analytic and numerical techniques for solving PDEs have proved unsuccessful,
approximation schemes have been introduced.  A particularly clear
and powerful approximation scheme for the low-energy dynamics of solitons
arising from Bogomol'nyi equations has been developed by N. Manton: the
geodesic approximation [1].  The geodesic approximation has been very
effective in describing the low-energy dynamics of BPS magnetic monopoles
and dyons [2-9], abelian vortices [10,11], solitons in
$\sigma$-models [12], Skyrme-like solitons in (2+1)-dimensions [13], and
Chern-Simons solitons [14].  However, we are only beginning to understand
the classical and quantum dynamics of these extended objects.

\smallskip
In this paper, which is divided into five sections, we study
the low-energy dynamics of BPS $SU(2)$ magnetic monopoles and dyons
encoded within the prototype of the standard electroweak model:
Yang-Mills-Higgs theory.  We present numerical results produced using
the geodesic approximation, which simulate the dynamics of two
low-energy BPS magnetic monopoles (or, dyons) with gauge group $SU(2)$
broken to $U(1)$ outside the magnetic monopole by an adjoint Higgs field.
We note that the representation of the Higgs field is not that used
by the standard model, and therefore our results do not comment directly on
the standard model.  In section one, we briefly state Manton's geodesic
approximation as it applies to BPS magnetic monopole classical dynamics
and introduce a closely related and carefully studied dynamical system
given by a Taub-NUT metric [3].  In section two we numerically simulate the
dynamics defined in section one to obtain a numerical approximation to the
classical differential scattering cross-section of BPS $SU(2)$ magnetic
monopoles.  In section three we are able to deduce some properties of the
magnetic monopole quantum differential cross-section by studying more
carefully the classical magnetic monopole cross-section and comparing it
to the classical and quantum Taub-NUT differential scattering
cross-sections.  In section four we widen our scope to include BPS magnetic
monopoles with an electric charge (dyons).  Our numerical work here examines
how the dynamics, both scattering and bounded, change qualitatively as we
vary initial conditions in phase space.  In particular, we look for initial
conditions in phase space which lead to either bounded or semi-bounded
motion.  Initial conditions leading to bounded BPS magnetic monopole dynamics
have been identified before [6,8], however, in those works the motion for
all time $t$ is required to remain close to spatial infinity.  We search
for initial conditions which give numerically stable, (semi-) bounded
motion over a long but finite time interval, that are not of the type
discussed in [6,8].  We also attempt to identify regions in phase space
which appear to be responsible for the non-integrable behaviour in the
dynamical system observed numerically in [8], and approximate slices of the
stable and unstable manifolds in phase space as proposed in [18].  The
study in section four completes the work begun in [4].  Section five is
a conclusion.

\bigskip
\noindent
{\bf 1. Magnetic monopole dynamics}
\medskip
Let ${\cal A}$ be the space of all static, $SU(2)$, finite-energy vector
potentials and Higgs fields, $(A,\Phi)\in{\cal A}$, on ${\bf R}^3$ in the
PS limit of an $SU(2)$ Yang-Mills-Higgs theory:
$${\cal E}= \int_{\bf R^3} d^3x\bigg(-{1\over 4}F_{ij}^aF^{ij}_a+
{1\over 2}(D_i\Phi)^a(D^i\Phi)_a\bigg).$$
The space of all smooth, gauge inequivalent, finite-energy field
configurations $(A,\Phi)$ forms an infinite dimensional
quotient subspace of ${\cal A}$, denoted by ${\cal C}={\cal A}/{\cal G}$.
Within ${\cal C}$ there exists a minimum energy submanifold $M_k$ given
by the smooth solutions---the BPS magnetic monopoles---to the
Bogomol'nyi equations, $B_i\equiv{1\over 2}\epsilon_{ijk}F_{jk}=D_i\Phi$.
The moduli space $M_k$ contains all BPS $SU(2)$ magnetic monopole of
magnetic charge $k$, where $k\equiv\int B_i^a(D\Phi)^a d^3x$.
Now let $k=2$, and consider $M_2$.  By factoring out translations in
Euclidean 3-space and an overall phase factor, $M_2$ reduces to the
quotient manifold $M^o_2$.  $M^o_2$ is of real dimension four.
The geodesic approximation states that the low-energy dynamics of
$k=2$ magnetic monopoles can be well-approximated by the geodesic
motion on $M^o_2$ [6].  The metric which defines the geodesics on the
moduli space $M^o_2$ was found by M. Atiyah and N. Hitchin by making use
of the natural hyper-K\"ahler structure which exists on $M^o_2$, and the
fact that in four dimensions a hyper-K\"ahler metric is equivalent to a
self-dual Einstein metric [2].

\smallskip
The metric on $M^o_2$ has been used to study $k=2$ BPS magnetic monopole
dynamics [2-9].  The moduli space $M^o_2$ can be parameterized by the
four coordinates: $r$, $\phi$, $\theta$, and $\psi$.  In these coordinates
the metric on $M^o_2$ can be written as
$$ds^2=f(r)^2\,dr^2+a(r)^2\,\sigma_1^2+b(r)^2\,\sigma_2^2+c(r)^2\,\sigma_3^2,$$
where $f$ is some function of $r$, and $\sigma_1$, $\sigma_2$ and $\sigma_3$
are the Euler differentials about the `body fixed' axes (the standard basis on
$so(3)^*$).  Imposing the self-dual Einstein condition on this metric, a system
of differential equations in the components of the metric is obtained [15,2].
The resulting equations are of two types: the equations describe a Taub-NUT
metric [15], or, they describe a metric first identified by Atiyah and
Hitchin [2].  In the case at hand, the Atiyah-Hitchin metric is the one
of interest and its metric components are given by the ordinary differential
equations [2]:
$${2bc\over f}{da\over dr}=(b-c)^2-a^2\ \ \ \ (cyclic),\eqno{(1)}$$
where `cyclic' means cyclic in $a$, $b$, and $c$.  For $f=abc$ [2] and
$f=-b/r$ [3] these equations have been analytically solved.  Below we
shall take $f=-b/r$.  Now the geodesic equations are easily computed.
$$\eqalign {{dM_1\over dt}&=\bigg({1\over b^2}-{1\over c^2}\bigg)M_2M_3
\ \ \ \ (cyclic),\cr
{d^2r\over dt^2}=-{1\over f}{df\over dr}\bigg({dr\over dt}\bigg)^2&+
{1\over f^2}\bigg({1\over a^3}{da\over dr}M_1^2+{1\over b^3}{db\over dr}M_2^2
+{1\over c^3}{dc\over dr}M_3^2\bigg).}\eqno{(2)}$$
For convenience $M_1\equiv a^2\sigma_1/dt$, $M_2\equiv b^2\sigma_2/dt$,
and $M_3\equiv c^2\sigma_3/dt$, and cyclic here means in $M_1$, $M_2$,
and $M_3$ and in $a$, $b$, and $c$ [2].  When the $k=2$ monopole solution
is extended asymptotically far in ${\bf R}^3$, the $k=2$ monopole can be
interpreted as two asymptotically separated $k=1$ monopoles.  For a
well-separated $k=2$ monopole, the expression $\int_\pi^r(b/r)\,dr$ can be
interpreted as the radial separation between the two $k=1$ monopoles;
$\phi$ and $\theta$ are the spherical polar coordinates which give the
angular position of the monopoles within ${\bf R}^3$; and $\psi$ is an
internal coordinate [2].  Equations (1) and (2) define the low-energy
dynamics of two BPS $SU(2)$ magnetic monopoles in a space free of other
particles and external fields.  The equations in (2) are a generalisation of
the Euler-Poinsot equations for a rigid body; in this case, the body is
not rigid and the moments of inertia can change with monopole separation.
With the coordinates we have used, there exists a coordinate singularity
at $r=\pi$ (a `bolt').

\smallskip
Numerical integration of the ordinary differential equations in (1) and
(2) indicate that $k>1$ $SU(2)$ BPS magnetic monopole dynamics is
non-integrable [4,8].  This appears to be due to the lack of relative
electric charge conservation, presumably analogous to the exchange of
charged gauge bosons in weak interactions.  In view of the
non-integrability of the dynamical system, it would be surprising if
an analytic expression for the BPS $SU(2)$ magnetic monopole differential
scattering cross-section were to exist.  At present no expression for
the cross-section is known.  We shall therefore determine the
differential scattering cross-section numerically, by modeling the
experimental procedure.  We do this in the next section.  We note
that all numerical experiments in this paper are designed so that the
data is taken only when the monopoles are well-separated, so that
interpretation of the data is understood physically.

\smallskip
Before we turn to the classical magnetic monopole differential
cross-section, we look at an interesting truncation of the Atiyah-Hitchin
metric discovered in [3] which {\it is} Liouville integrable.  When the
Atiyah-Hitchin metric is expanded in large $r$ and exponentially small terms
in the metric coefficients are dropped, the truncated metric written in the
coordinates $r$, $\phi$, $\theta$, and $\psi$ is
$$ds^2=\big(1-{2\over r}\big)\,dr^2+r^2\big(1-{2\over r}\big)(d\theta^2+
\sin^2\theta\,d\phi^2\big)+{4r\over r-2}(d\psi+\cos\theta\,d\phi)^2.
\eqno{(3)}$$
The metric in (3) is a Taub-NUT metric with a negative mass parameter
and has a coordinate singularity at $r=2$.  Asymptotically, BPS
magnetic monopole dynamics behaves like Taub-NUT geodesic dynamics.
The geodesic equations are straightforward to compute and so will not be
explicitly stated.  Unlike two BPS magnetic monopole dynamics, Taub-NUT
geodesic dynamics can be shown to be Liouville integrable.  In the
Hamiltonian setting there are four constants of motion for the eight
dimensional phase space: the angular momentum, the generalized momentum of
the cyclic variable $\phi$, the electric charge difference between the
particles given by $p_\psi=c^2({\dot\phi}\cos\theta+\dot\psi)$, and the
energy.  As a result, the classical differential scattering cross-section
can be found [3]:
$$\bigg({d\sigma\over d\Omega}\bigg)_{TN}^C={1\over4}\big( g^2+\big({q\over v}
\big)^2\big)^2\csc^4{\theta\over 2},\eqno{(4)}$$
where $g=1$, $q$ is half the relative electric charge difference, $v$ is the
impact velocity, $\theta$ is the scattering angle in the centre-of-mass
frame, and the reduced mass of the magnetic monopole system is taken to be 1.

\smallskip
The differential cross-section (4) is a generalisation to the Rutherford
differential cross-section in electrodynamics (formally we let $g\to 0$).
However, as a generalisation of electrodynamics it differs from that
proposed by Poincar\'e and Dirac (see [19], for example), which has a
differential cross-section of the form:
$$\bigg({d\sigma\over d\Omega}\bigg)_{EM}\simeq{1\over4}\bigg(
{g^2+q^2\over v^2}\bigg)^2\bigg({\theta\over 2}\bigg)^{-4},$$
for small scattering angles.  Of course, the Taub-NUT dynamics differs
from the conventional generalisation of electrodynamics even at small
scattering angles because it possesses a (massless) Higgs field which in
the static case with $q=0$ exactly cancels the Coulombic magnetic
interaction of conventional magnetic monopoles.  The structural similarity
of Taub-NUT dynamics to Coulomb dynamics and the Liouville integrability of
Taub-NUT, suggests that Taub-NUT dynamics is of independent interest.
Classical and quantum Taub-NUT dynamics has been thoroughly studied by
Gibbons and Manton [3].  The Taub-NUT cross-sections will be useful
in this paper.
\vfill\break\eject

\noindent
{\bf 2. The classical differential scattering cross-section}
\medskip
In this section we shall use the geodesic approximation to compute
the differential scattering cross-section of pure $SU(2)$ magnetic
monopole dynamics in the centre-of-mass frame.
The plots in figures 1 and 2 arise from the numerical simulation of
over 32 000 magnetic monopole scattering experiments using 16K parallel
processors on a Connection Machine.

\smallskip
To produce our results the radial distance between each pair of magnetic
monopoles is fixed with one, single radial separation, $R_0>>0$.  The
value of the impact parameter for each pair of magnetic monopole is chosen
by a random number generator (based on cellular automata) uniformly
distributed between 0 and a maximum value for the impact parameter, $I$,
which we are free to choose.  The scattering plane for each
pair of magnetic monopole is set to $\theta=\pi/2$ and $\dot\theta=0$,
using the rotational invariance of two magnetic monopole dynamics.  Although
rotations do not act trivially on the relative coordinate, $\psi$, we may
choose a scattering plane with impunity because $\psi$ is set, like $I$,
using a random number generator.  $\psi$ takes its values between $0$ and
$2\pi$ for each two particle interaction.  The impact velocity for all
pairs of magnetic monopole is set to -1.0, and we compute the radial and
angular velocity components for each pair using the value of the impact
parameter.  The magnetic monopoles are initially uncharged: $\dot\psi=0$.

\smallskip
With the initial conditions now completely described, the monopoles are
scattered against each other by numerically integrating the Atiyah-Hitchin
geodesic equations (1) and (2) with $f=-b/r$ using a fourth-order
Runge-Kutta scheme.
As a test of our numerical procedure below, we also study numerically
the integrable Taub-NUT dynamics.  The numerical integration of both sets
of geodesic equations (those using the Atiyah-Hitchin metric (1) and (2),
and those equations arising from the truncated Taub-NUT metric in (3)) have
been extensive tested.  The known constants of motion in each case are
respected within numerical accuracy, and for a number of values in allowable
error for each step taken in the Runge-Kutta scheme.  For the Atiyah-Hitchin
metric the scattering dynamics is in agreement with the analytically known
behaviour for $\psi=0$ and $\psi=\pi/2$ where explicit computation has been
possible [3,5], and the bounded dynamics is in agreement with the
predictions of secular perturbation theory when the equations are
interpreted as a near-integrable Hamiltonian system [8].

\smallskip
After numerically integrating the equations, the scattering angles are
sorted into $N_{BIN}\in{\bf Z}$ equally sized bins centered round
$\theta=\pi/2$.  To compute the differential cross-section, let
$N(\Delta\Omega)$ be the number of magnetic monopoles scattering into the
solid angle $\Delta\Omega$, and $N_T$ the total number of scattering
experiments.  This takes into account the indistinguishability of magnetic
monopoles.  The differential cross-section is given by
$${d\sigma\over d\Omega}=
\mathop{\mathop{\lim_{I\to\infty}}_{N_T\to\infty}}_{\Delta\Omega
\to 0}{\pi I^2\over\Delta\Omega}{N(\Delta\Omega)\over N_T}.
\eqno{(5)}$$
Our numerical work computes (5) bearing in mind that we can only
approximate the limits.  Therefore before we apply this procedure to the
magnetic monopole, we test the reliability of our numerical scheme for
differential cross-sections by comparing the analytical and numerical
classical differential cross-sections for the completely integrable
Taub-NUT dynamical system.  In Fig. 1, we see that the numerical code
applied to Taub-NUT dynamics gives good agreement with the analytical
differential cross-section (4) (solid curve) over a wide range of angles.
The error bars are explained below (see Error Analysis).  Fig. 1 is
produced by setting $I=3.0$ and $N_{BIN}=79$.  Only 59 data points are
plotted.  The 20 points excluded are on either the far left or far right
of the plotted points, and their values are meaningless artifacts of the
finite impact parameter.  The value of $N_{BIN}$ is made as large as
possible, until statistical fluctuations prevent us from going any further.
The small but significant discrepancy between the analytic and numeric
cross-sections is due to the finite values given to $I$, $N_T$, and
$N_{BIN}$.  To improve the numerical cross-section a larger value of $I$
is necessary.  But, in order to maintain the statistics we would need to
increase the number of particle experiments.  Numerical experiments appear
to confirm the expected improvement, but more resources from the
Connection Machine (which are unavailable at present) would be required to
produce figures with as many data points as figures 1 and 2 without
statistical fluctuations.

\smallskip
The best results we can get for the magnetic monopole differential
cross-section are reproduced in Fig. 2 with $I=3.0$ and $N_{BIN}=29$.
For $N_{BIN}>29$ statistical fluctuations become apparent.
We have fit an eighth order polynomial (the solid curve) to the 21 plotted
data points.  This is only an approximation to the differential
cross-section due to the finite values given to $I$, $N_T$, and
$N_{BIN}$, as was the case with the Taub-NUT cross-section.  However,
the general shape, particularly the hump at $\theta=\pi/2$,
can be viewed as accurate, because increasing $I$ and $N_T$ will
have the effect of push up the sides (the small angle scattering), and
deepening the double well pattern in the cross-section.
Moreover, increasing $N_{BIN}$ reduces the amount of averaging in the
cross-section, further exposing the structure in Fig. 2 before
fluctuations become important.
The most striking feature of Fig. 2 is that it does not look like a
{\it classical} bosonic differential cross-section.  Rather, it looks
like a {\it quantum} bosonic differential cross-section.  We are able
to explore this in more detail in the next section.

\bigskip
\noindent
{\it Error Analysis.}
\medskip
We discuss the two sources of error which are used to compute the error
bars above: bolt error and numerical error.

\smallskip
The first source of error---bolt error---occurs because our
numerical dynamics become unreliable near the bolt.  One solution to
this problem is to remove and identify any experiment which comes too
close (say within 0.01) to the bolt, and incorporate the number
of bolts into the error.  The number of experiments we remove as a
result of this strategy is less than 0.6\% of the total number
of experiments for $I=1.0$, and approximately 0.1\% for $I=3.0$.
We note that the number of bolt events decreases, as we
increase $I$ to better approximate the cross-section.  Although there
are relatively few bolts in our experiments, one might be inclined to
approach the bolts differently: to remove them entirely by introducing
coordinate patches.  We could then integrate through the singularities.
With this approach one would find that the coordinate transformations are
difficult to verify, and, worse, the reader would be obliged to take on
faith that the coordinate transformations are coded correctly.  Moreover,
one would need to use approximations to the metric coefficients near the
bolt which would undermine the results.  Due to the low number of
bolt events such steps are not necessary, and we can afford to make
our procedure more transparent by returning to our first approach:
clearly identify those scattering experiments with dynamics which come
too close to the bolt, and incorporate them into our error analysis.  The
advantage to this treatment of the region around the bolt is that these
events are clearly seen and their relative impact on our results easily
judged.

\smallskip
We turn now to the second source of error---the numerical error.
To account for the numerical error we allow monopole events detected in
any bin to be reclassified as events in an adjacent bin.  With a numerical
error of $10^{-5}$ per step, we estimate that at worst 1.5\% of the magnetic
monopoles in bins adjacent to any particular bin, $B_k$, could have scattered
into $B_k$.

\smallskip
Errors bars are added to Figs. 1 and 2 by assuming that the bolts and
the reclassifiable events scatter in the most detrimental manner to our
differential cross-sections.  Explicitly for Fig. 1, we
include all the bolts and reclassifiable events into each particular bin under
analysis, $B_k$, to get an upper bound on the differential cross-section at
that bin.  For the lower bound on the differential cross-section at the bin
$B_k$ we include no bolts and remove reclassifiable events in the bin
$B_k$ (3\% of $B_k$).  This is repeated for Fig. 2.  There is little
variation in our results when the number of monopoles studied is
reduced by a factor of one half, or, if the permitted numerical error per
step is changed to $10^{-4}$ or $10^{-6}$.

\bigskip
\noindent
{\bf 3. The quantum differential scattering cross-section}
\medskip
In this section we explore properties of the quantum magnetic monopole
differential cross-section in the centre-of-mass frame.  We begin by
studying the quotient of the Taub-NUT classical differential cross-section
by the $SU(2)$ magnetic monopole differential cross-section given by the
Atiyah-Hitchin metric:
$${d\sigma_{TN}/d\Omega\over d\sigma_{AH}/d\Omega}=
{dN_{TN}\over dN_{AH}},\eqno{(6)}$$
in the centre-of-mass frame.  Fig. 3 plots the quotient in (6) using the
numerically computed classical cross-sections, magnetic monopole and
Taub-NUT, of the previous section both with 29 bins.  In particular, as
we can see from fig. 3 the ratio in (6) is approximately $1/2$ at
$\theta=\pi/2$.  Unlike the differential cross-sections in the previous
section, the ratio of cross-sections does not appear to be as sensitive to
the maximum impact parameter, $I$.  This is due to the special relationship
between the Atiyah-Hitchin metric and the Taub-NUT metric.
For magnetic monopoles with a large impact parameter, the dynamics
are well-described by the Taub-NUT metric.  Therefore by decreasing
$I$ we remove approximately the same number of small angle scattering
experiments in both Atiyah-Hitchin and Taub-NUT dynamics, and replace them
uniformly with impact parameters which give larger scattering angles.  This
improves the statistics without significantly changing the ratio
$dN_{AH}/dN_{TN}$.  By decreasing $I$, however, we also shorten the range
of angles around $\theta=\pi/2$ for which the computed ratio is accurate.
In Table 1 we use the relationship between magnetic monopole scattering
and Taub-NUT scattering to further test the ratio in Fig. 3 at $\pi/2$
for other values of maximum impact parameter, $I$.  Errors have been
computed for the ratios contained in Table 1, and are included there.
How these errors are computed is explained in the Error Analysis at
the end of this section.  Table 1 supports the claim that at
low-energies the BPS magnetic monopole classical cross-section is
approximately twice that of the classical Taub-NUT cross-section at
$\theta=\pi/2$.

\smallskip
Due to the integrability of classical Taub-NUT dynamics, quantum
Taub-NUT dynamics is also well-understood [3].  The Taub-NUT metric is
used to define the covariant Laplacian in the Schr\"odinger equation,
and the quantum scattering problem can be solved exactly in parabolic
coordinates.  The quantum Taub-NUT differential cross-section is found
to be [3]:
$$\bigg({d\sigma\over d\Omega}\bigg)_{TN}^Q=
{1\over 4}\bigg(\csc^4{\theta\over 2}+\sec^4{\theta\over 2}\bigg)+
{1\over 2}\sec^2{\theta\over 2}\csc^2{\theta\over 2},\eqno{(7)}$$
taking into account the indistinguishability of the particles.
As we would expect at $\theta=\pi/2$, the classical Taub-NUT
cross-section (4) and the quantum Taub-NUT cross-section differ
by a factor of exactly two, due to the quantum mechanical exchange
effects for identical bosons.  All of this, of course, is
elementary.  But in view of Fig. 3 and Table 1, we must conclude that
classical magnetic monopole scattering is approximately equal to quantum
Taub-NUT at $\theta=\pi/2$.  The significance of this becomes
clearer when we make use of Schroers recent study on low-energy
quantum BPS $SU(2)$ magnetic monopole scattering [9].  As in [3],
the Atiyah-Hitchin metric is used by Schroers to define the covariant
Laplacian in the Schr\"odinger equation.  A partial wave analysis indicates
that at sufficiently low energies where only the s-wave significantly
contributes to the scattering amplitude, the quantum BPS magnetic monopole
scattering amplitude is, to within numerical accuracy, equal to the quantum
Taub-NUT scattering amplitude [9].  We have repeated Schroers numerical
work and can confirm his result.  Since the quantum Taub-NUT cross-section
is a good approximation to the quantum magnetic monopole cross-section [9],
Fig. 3 and Table 1 now suggest that the BPS $SU(2)$ magnetic monopole
differential cross-section is about the same at $\theta=\pi/2$ in both
the classical and quantum theories in the low-energy limit!  This is
surprising because in both the BPS magnetic monopole theory and the
Taub-NUT theory the particles are identical bosons; therefore
on very general quantum mechanical grounds we would expect the classical
and quantum cross-sections to differ by a factor of two at the scattering
angle $\theta=\pi/2$ (cf. our Fig. 3 with Fig. 2 in [16]) [16].  We have
already observed above a factor of two at $\theta=\pi/2$ in classical and
quantum Taub-NUT, but a factor of two can be {\it excluded} in our
numerical experiments for BPS magnetic monopoles.  We now summarize our
low-energy observations:
$$\bigg({d\sigma\over d\Omega}\bigg)_{AH}^C\bigg\vert_{\theta=
{\pi\over 2}}\sim 2\times\bigg({d\sigma\over d\Omega}\bigg)_{TN}^C
\bigg\vert_{\theta={\pi\over 2}}\sim\bigg({d\sigma\over d\Omega}
\bigg)_{TN}^Q\bigg\vert_{\theta={\pi\over 2}}\sim\bigg({d\sigma\over
d\Omega}\bigg)_{AH}^Q\bigg\vert_{\theta={\pi\over 2}}.$$
The conclusion lies at the ends of the chain.

\smallskip
The missing quantum mechanical exchange effects for indistinguishable
BPS $SU(2)$ magnetic monopoles is at first disturbing.  Why is the factor
of two present in Taub-NUT dynamics and not in BPS $SU(2)$ magnetic
monopole dynamics?  The origin of the discrepancy lies in the classical
configuration spaces.  We first look at classical Taub-NUT dynamics.
Classical Taub-NUT dynamics is scale-free, so that the finite size of
the magnetic monopoles becomes unimportant.  Therefore Taub-NUT
dynamics may be thought of as the dynamics of magnetic monopole
point-particles [3].  If we denote the one point-particle magnetic
monopole configuration space by $M_1$, then the two point-particle
magnetic monopole configuration space is the symmetric product space
$S^2(M_1)$, where we have taken into account the indistinguishability
of the $k=1$ magnetic monopoles.  But the actual BPS $SU(2)$ magnetic
monopole symmetrised configuration space is $M_2$ (we do not fix the
centre-of-mass or the overall phase).  $M_2$ is very different from
$S^2(M_1)$.  The most noticeable difference between the two is that
$S^2(M_1)$ possesses singularities where two point-particles coincide,
while $M_2$ has no real singularities [2].
The lack of singularities in $M_2$ boils down to the fact that BPS
$SU(2)$ magnetic monopoles cannot be viewed as point-particles.  This
expresses itself quantum mechanically as well.  Symmetrisation in
point-particle quantum mechanics can be formulated in two ways: one can
define a symmetrised wavefunction on the unsymmetrised configuration space
$M_1\times M_1$ (the conventional approach); or, one can define a
wavefunction on the symmetrised configuration space $S^2(M_1)$.  For
point-particles the approaches are equivalent [17].  For BPS $SU(2)$
magnetic monopoles, however, we {\it must} take the second approach,
because $M_2$ cannot be expressed as the symmetric product of two
one-particle configuration spaces.  Therefore there need not be quantum
mechanical exchange effects of the type envisioned by Mott.
This conclusion is compatible with our numerical results.

\bigskip
\noindent
{\it Error Analysis.}
\medskip
Our treatment of error in this section is very similar to what
we did in the previous section.  In Fig. 3 we computed the error bars
by adding the Taub-NUT bolts and reclassifiables to the numerator
at each bin, and subtracting Atiyah-Hitchin reclassifiables from the
denominator at each bin.  This gives the upper bound.  Subtracting
Taub-NUT reclassifiables from the numerator at each bin, and adding
Atiyah-Hitchin bolts and reclassifiables to the denominator at each bin
gives the lower bound.  In Table 1 the same computation is performed
on only one bin, the bin containing $\theta=\pi/2$, but the computations
are done for a number of maximum impact parameter values, $I$.
The highest and lowest value for the ratio has been include in Table 1.

\bigskip
\noindent
{\bf 4. BPS magnetic monopole escape plots}
\medskip
While in the previous section we studied the scattering of BPS
magnetic monopoles within $SU(2)$ Yang-Mills-Higgs gauge theory, this
section attempts to understand both the scattering and bounded motion
of BPS dyons (magnetic monopoles with a relative electric charge difference).
We examine how the initial conditions leading to each type of motion coexist
in phase space.  The existence of bounded BPS dyon motion was argued in [6,8]
by interpreting the low-energy BPS dyon dynamical system modeled by the
Atiyah-Hitchin metric (2) as a perturbation of the Liouville integrable
Taub-NUT dynamics given by the metric (3).  This required that the dynamics
remain sufficiently close to the asymptotic region throughout its motion.  We
are unaware of any initial conditions which give stable bounded motion that
are not of the type discussed in [6,8], although we note that an {\it unstable}
bounded motion outside of the asymptotic region has been found [5].
In [4] it was proposed that the dynamics of two BPS dyons could be described
using numerically generated `Julia sets' (we call these `escape plots' below).
To speed-up the numerical computation on a serial computer in [4] it was
necessary to introduce an artificial Poincar\'e plane, acting as a rapid
timing device.  Despite this shortcoming a definite structure was detected,
suggesting that the Poincar\'e plane should be removed when possible
and the true global dynamical structure of the phase space exposed.  This
is possible with massively parallel computation.

\smallskip
We propose to construct and display escape plots for dyon dynamics
in this section.  Our purpose is to identify initial conditions which appear
to be bounded and numerically stable, but which do not remain in the
asymptotic region studied in [6,8].  In addition, the escape plots shall
suggest that initial conditions exist which are only semi-bounded, so that
scattering initial conditions can be entrapped.  In [8] the Hamiltonian
formulation of BPS magnetic monopole scattering was studied, and numerical
modeling of the Hamiltonian system indicated that the BPS $SU(2)$ magnetic
monopole two body problem is not integrable.  Structure within the escape
plots in this section gives further evidence for the non-integrability of
magnetic monopole dynamics.  As explained in [18,4], the escape plots can
also give an approximation to the stable and unstable manifolds in phase
space.

\smallskip
We shall first discuss the initial conditions.  Magnetic monopole
(dyon) dynamics can be rewritten in the Hamiltonian formulation.  The
conjugate momenta are found to be
$$\eqalign {&p_r=f^2v,\ \ \ p_\psi=M_3,\cr
&p_\theta=M_2\cos\psi-M_1\sin\psi,\cr
&p_\phi=M_1\cos\psi\sin\theta+M_2\sin\psi\sin\theta+M_3\cos\theta,}$$
where we have used the notation of section one.
Asymptotically, $p_\psi$ can be seen to be proportional to the relative
electric charge of the monopole system and is a conserved quantity [3].
But in general $p_\psi$ is not a conserved quantity.  To construct
the escape plots for two dyon dynamics, we shall reduce the size of the
eight dimensional phase space by restricting to the level set of the
conserved total angular momentum, $M^2$.  In addition, using the $SO(3)$
invariance of the metric on $M^o_2$, we may set the initial scattering
plane to that with
$$\phi=0,\ \ \theta=\pi /2,\ \ \ and\ \ \dot\theta=0.$$
At the initial point, therefore, the total angular momentum
$$M^2=p_\phi^2+p_\theta^2+p_\psi^2.$$
We also take advantage of the scale invariance of the geodesic
motion by specifying that the angular speed in Euclidean
three-space is minus one,
$$r\dot\phi=-1,$$
because only the direction of the tangent vector to the level set is
needed to trace out the motion.  Organised in this way, $M^2$, $\psi$,
$p_r$, and $r$ parametrise the initial conditions.  We shall now enclose the
dyons in a large spatial sphere, where the centre-of-mass is located at the
centre of the sphere.  The initial conditions with dyon separation within the
sphere form a submanifold of the phase space.  It is these initial conditions
that we study in our escape plots below.

\smallskip
We have found that the most attractive results are achieved when the
initial conditions are taken from the level surface, $S$,
determined by setting $M^2$ and $p_r$ to constants, for then the escape
plots are parameterized by $\psi$ and $r$ as polar coordinates.
Note that $p_r$ is not a constant of the motion.  There is much to be
gained by rescaling $r$, as well.  Therefore, we shall parameterized the
escape plot with the polar coordinates $(\rho(r),\psi)$, with
$$r=2K(\sin(\rho/2)),$$
where $K$ is the complete elliptic integral of the first kind.  This has
the effect of, one, bringing infinite dyon separation to $\rho=\pi$,
and, two, magnifying that part of the escape plot in which we are most
interested (those initial conditions {\it within} the sphere).  We now place
a fine two-dimensional grid on $S$, where each grid point determines a full
set of initial conditions for a two dyon dynamical experiment.  For each
set of initial conditions, equivalently, each grid point, we numerically
integrate (Runge-Kutta, again) the equations of motion (1) and (2) over a
large interval of time restricted by the accumulated error.  An initial
condition and a copy of the equations of motion are installed on each
processing unit of the parallel processing computer.  If the dyons reach the
large sphere (of `detectors'), then we note the amount of time taken to get
to the sphere.  We shall plot this information and call it an escape plot.
We have examined escape plots for other choices of level surface (e.g., when
$p_r$ and $r$ parameterized the escape plot), and the structure is found to
be very similar but not as symmetric.  Furthermore, changes in the
integration time, in the radius of the spatial sphere, and in the allowable
numerical error do not significantly change our results.

\smallskip
The escape plots in plates 1 and 2 are constructed by choosing the level
surface, $S$, given by $M^2=400.0$ and $p_r=-1.25$.  With $p_r=-1.25$
the magnetic monopoles are directed out of the asymptotic region, so that
the analysis in [6,8] is no longer valid.  We colour a grid point
on $S$ according to its `escape time': the time taken to reach a large
surrounding sphere.  Dark blue indicates that the initial condition reached
the large sphere relatively quickly; light blue initial conditions are a bit
slower in reaching the sphere; light green initial conditions just barely
reach the sphere in the integration time; and dark green initial conditions do
not reach the sphere at all.  This is similar to the Julia set in the
theory of iterated functions defined on the complex plane.  We
shall run the initial conditions for both increasing time (plate 1) and
decreasing time (plate 2).  The plotting area is circular and the boundaries
correspond to $\rho=2.5$ in plate 1, and $\rho=2.2$ in plate 2.  Nb., the
centre of both plots is $\rho=\pi$ ($r=\infty$!).  The structure in plate 1
fits into the central spiral of plate 2.  Note also that when a pixel
corresponds to an initial condition with a dyon separation comparable to, or,
greater than the radius of the surrounding sphere, the escape information is
sensitive to the radius of the sphere, and therefore not of interest to us.
Therefore the ring-like structure in a small region round the origin,
$r=\infty$, should be ignored.

\smallskip
Semi-bounded initial conditions exist: dark green in one time direction,
and dark blue in the other.  This leads to the entrapment of scattering
initial conditions.  Physically, dyons coming from infinity can acquire a
relative electric charge difference in their interaction with each other
to form a bound state.  This does not violate Liouville's theorem (that
Hamiltonian dynamics preserves phase space volume) because there are the
extra coordinates $\psi$ and $p_\psi$ which can absorb volume.  Furthermore,
since there appears to be no reason to favour one time direction over the
other, we may superimpose plates 1 and 2 on the same scale.  Intersections
of the dark green in the increasing time direction (plate 1), and dark green
in the decreasing time direction (plate 2) indicate that there are initial
conditions leading to motion bounded for all time.  Moreover, numerical
error accumulated over the long integration time would suggest that the
bounded dynamics which we observe in the escape plots are numerically
stable (see Error Analysis below).
We call a numerically stable motion bounded in both time directions
a `bound state'.  As the maximum integrating time is increased, the dark
green regions thin, suggesting that semi-bounded initial conditions in each
time direction are probably one dimensional.  If so, the intersections which
give bound states are isolated and give a discrete energy spectrum for dyons.
The energies can be computed from
$$\eqalign {E(r,\psi,p_r,M^2)={p_r^2\over f(r)^2}&+{a(r)^2\cos^2\psi\over r^2}
+{b(r)^2\sin^2\psi\over r^2}\cr
&+{1\over c(r)^2}\bigg(M^2-{a(r)^4
\cos^2\psi\over r^2}-{b(r)^4\sin^2\psi\over r^2}\bigg).}$$
We are unable to compute the bound state energies at present for
somewhat technical reasons to due with the compiler.

\smallskip
The dark green regions may contain stable and unstable fixed points---when
they do, the dark green regions are approximations to stable and
unstable manifolds, as explain in [18,4].  The spiraling of colours
into points or small limit cycles, and, the presence of `stability filaments'
are significant as well.  In regions where colours spiral we expect a
sensitivity to initial conditions; these regions appear to be responsible
for stochastic behaviour and are further evidence for the non-integrability
of the two BPS magnetic monopole dynamical system.  Stability filaments have
also been observed in the Cremona map, an area preserving twist map [18].
The Cremona map has been argued to contain the generic structure of
Hamiltonian dynamics, and thereby to be of physical importance [11].  The
structures in plates 1 and 2 would seem to support this view.  It is probably
in a detailed analysis of the Cremona map and its properties, that the
dynamical structure of two BPS magnetic monopole escape plots will be more
fully appreciated.  The numerical work in [11] suggests that the stability
filaments often coincide with the `characteristics' of periodic points.
Stability filaments in plates 1 and 2 may be indirect evidence for stable
periodic orbits which are not in the near-asymptotic region (cf., [5,8]).

\bigskip
\noindent
{\it Error Analysis.}
\medskip
At this point it is convenient to mention an ommission in the escape plots;
initial conditions which come too close to the bolt have not been
explicitly represented in these plots.  The number of near bolt events
was less than 0.5\% of the total number of initial conditions examined.
The initial conditions in plates 1 and 2 which led to dynamics that came
too close to the bolt formed two thin curves (a few pixels wide).
In plate 1, one of the curves leaves the circular boundary at about
`10 o'clock' and spirals into the closest limit point (or, cycle); and
the other leaves the circular boundary at about `4 o'clock' and spirals
into its closest limit point (or, cycle).
A similar pair of curves is found in plate 2.
The events which come too close to the bolt are contained {\it entirely}
within the dark blue regions.  Their location and their insensitivity
to changes in the bolt condition suggest that the bolts do not effect
the general structure of the escape plots.

\smallskip
The escape plots, including the placement of the bolts, are also robust
to changes in the allowable error, changes in the radius of the
surrounding sphere, and increases in the total integration time.

\bigskip
\noindent
{\bf 5. Conclusion}
\medskip
Our experiments were limited by the number of processors we could get
consistent access to---with an entire CM2 at our disposal we could have
increased the number of magnetic monopole experiments by a factor of four,
and would have improved our results in sections 2 and 3 by increasing
the maximum impact parameter and the number of bins.  We believe, however,
that the essential features of the magnetic monopole differential
cross-section in Fig. 2 will be maintained with improvements to the
numerical method.  An increase in the number of processors would have
had very little effect on the escape plots in plates 1 and 2, although
it would be useful in computing the fractal dimension.  Therefore,
based on our numerical work our conclusion is as follows.
We have shown that features associated with the low-energy quantum
dynamics of point-particles are present in the classical
dynamics of solitonic BPS $SU(2)$ magnetic monopoles.  One
would very much like to know if this behaviour is reproduced in
other solitonic systems which can not be modeled as point-particles.

\vskip 0.25in
\openup -2\jot
\medskip
\noindent
{\it Acknowledgements.} We are grateful for time on a Connection Machine
located at Connection Machines, Corp., Cambridge, Mass., and supported by
DARPA contract DACA76-88-C-0012.  The colour plots were produced on an
IRIS.  TR thanks the University of Arizona for their hospitality and
assistance, and R. Jackiw for useful suggestions.  This work benefits from
an NSERC research grant (OGP0105498,TR), and an NSERC undergraduate student
research award (DA).

\vfill\break\eject
\openup 2\jot
\vskip 0.5in
\centerline{{\bf References}}
\smallskip
\noindent
\item {[1]} N.S. Manton, {\it Phys. Lett.} B 110 (1982) p.54.
\smallskip
\noindent
\item {[2]} M. Atiyah, N. Hitchin, The geometry and dynamics of magnetic
monopoles (Princeton U.P., Princeton, NJ, 1988).
\smallskip
\noindent
\item {[3]} G.W. Gibbons, N.S. Manton, {\it Nucl. Phys.} B {\bf 274}
(1986) 183.
\smallskip
\noindent
\item {[4]} M. Temple-Raston, {\it Phys. Lett.} B 213 (1988) pp.168-71.
\smallskip
\noindent
\item {[5]} L. Bates, R. Montgomery, {\it Comm. Math. Phys.} {\bf 118}
(1988) p.635.
\smallskip
\noindent
\item {[6]} M.P. Wojtkowski, {\it Bull. Amer. Math. Soc.} 18 (1988) 179.
\smallskip
\noindent
\item {[7]} M. Temple-Raston, {\it Phys. Lett.} B 206 (1988) 503.
\smallskip
\noindent
\item {[8]} M. Temple-Raston, {\it Nucl. Phys.} B {\bf 313} (1989) 447-72.
\smallskip
\noindent
\item {[9]} B.J. Schoers, {\it Nucl. Phys.} B {\bf 367} (1991) 177-214.
\smallskip
\noindent
\item{[10]} T.M. Samols, {\it Comm. Math. Phys.} {\bf 145} (1992) 149-80.
\smallskip
\noindent
\item{[11]} P.J. Ruback, {\it Nucl. Phys.} B {\bf 296} (1988) 669.
\smallskip
\noindent
\item{[12]} R.A. Leese, {\it Nucl. Phys.} B {\bf 334} (1990) 33.
\smallskip
\noindent
\item{[13]} P.M. Sutcliffe, {\it Nonlinearity} {\bf 4} (1991) 1109.
\smallskip
\noindent
\item{[14]} Long Hua, Chihong Chou, MIT-CTP 2064, January 1992.
\smallskip
\noindent
\item {[15]} G.W. Gibbons, C. Pope, {\it Comm. Math. Phys.} {\bf 66}
(1979) 267-90.
\smallskip
\noindent
\item {[16]} N.F. Mott, {\it Proc. Roy. Soc. Lond} A{\bf 126} (1930)
pp.259-67.
\smallskip
\noindent
\item {[17]} J.M. Leinaas, J. Myrheim, {\it Il Nuovo Cimento}
{\bf 37}B (1977) 1-23.
\smallskip
\noindent
\item {[18]} E.C. V\'azquez, W.H. Jefferys, A. Sivaramakrishnan,
{\it Physica} D 29 (1987) 84.
\smallskip
\noindent
\item {[19]} J. Schwinger, K.A. Milton, W-Y Tsai, L.L. DeRand, Jr.,
D.C. Clark, {\it Ann. of Phys.} {\bf 101} (1976) pp.451-95.

\vfill\break\eject
\openup -2\jot
\vskip 1in
\centerline {\bf Figure Captions}
\vskip 0.25in
\noindent
{\bf Figure 1.} The classical differential cross-section for Taub-NUT
scattering.  The numerical cross-section is computed with $I=3.0$ and
$N_{BIN}=79$; 61 data points are plotted.  The solid curve is
the evaluation of the analytic cross-section.
\vskip 0.25in
\noindent
{\bf Figure 2.} The classical differential cross-section for $SU(2)$
BPS magnetic monopole scattering with $I=3.0$ and $N_{BIN}=29$.  An
eighth order polynomial fit to the 21 data points is given by the
solid curve.
\vskip 0.25in
\noindent
{\bf Figure 3.} The ratio of the classical Taub-NUT differential
cross-section to the classical $SU(2)$ BPS magnetic monopole
differential cross-section with $I=3.0$ and $N_{BIN}=29$.
\vskip 0.25in
\noindent
{\bf Table 1.} The ratio of the classical Taub-NUT differential
cross-section to the classical $SU(2)$ BPS magnetic monopole
differential cross-section at $\theta=\pi/2$ for $I=1.0,1.5,2.0,3.0$.
\vskip 0.25in
\noindent
{\bf Plate 1.} The escape plot for $M^2=400.0$ and $p_r=-1.25$ in the
increasing time direction.  We use the coordinates $\rho(r)$ and $\psi$ as
polar coordinates on the plot.
\vskip 0.25in
\noindent
{\bf Plate 2.} The escape plot for $M^2=400.0$ and $p_r=-1.25$ in the
decreasing time direction.  We use the coordinates $\rho(r)$ and $\psi$ as
polar coordinates on the plot.